\documentclass[
  amsmath,amssymb,
  aps,reprint,superscriptaddress,floatfix
]{revtex4-2}
\usepackage{graphicx}
\usepackage{dcolumn}
\usepackage{bm}
\usepackage{hyperref}
\usepackage{braket}
\usepackage{dsfont}
\usepackage{xcolor}

\begin{document}
\title{Robust Ion-Photon Entanglement via Polarization-to-Time-Bin Conversion}
\author{Ana Luiza Ferrari}
\email{linke@umd.edu, afr27@duke.edu}
\author{Denton Wu}
\author{Mika A. Zalewski}
\affiliation{Duke Quantum Center, Department of Electrical and Computer Engineering and Department of Physics, Duke University, Durham, NC 27708, USA}
\author{Norbert M. Linke}
\email{linke@umd.edu, afr27@duke.edu}
\affiliation{Duke Quantum Center, Department of Electrical and Computer Engineering and Department of Physics, Duke University, Durham, NC 27708, USA}
\affiliation{Joint Quantum Institute and Department of Physics, University of Maryland, College Park, MD 20742, USA}
\affiliation{National Quantum Laboratory (QLab), University of Maryland, College Park, MD 20742, USA}
\date{\today}
\begin{abstract}
Time-bin photonic qubits are well-suited for quantum network applications due to their robustness to polarization instability in fiber links and potential for heterogeneous networks. In this work, we implement the first entanglement-preserving polarization-to-time-bin conversion of a photon qubit in an entangled state with a matter qubit, together with the independent work of Haen \textit{et al.} \cite{haen2026telecom}. Photons initially generated with polarization encoding are converted to the time-bin basis through a polarization-discriminating asymmetric Mach--Zehnder interferometer. The photonic qubits are generated via the $1092$~nm transition of a $^{88}$Sr$^{+}$ ion. We measure state fidelity bounds of $0.906 \pm 0.011 \le \mathcal{F} \le 0.934\pm 0.011$, with conversion error $< 0.028$, and find this fidelity shows no statistically significant reduction under depolarizing noise, even at full depolarization strength.
\end{abstract}
\maketitle


\section{Introduction}

Quantum networking is a rapidly developing field with many technology goals. Beyond distributed quantum computing \cite{Chris_QN_original}, quantum networks enable distributed sensing \cite{proctor2018multiparameter, zhang2021distributed},  metrology \cite{nichol2022elementary}, and secure quantum communication \cite{gisin2002quantum, nadlinger2022experimental} applications, all of which require physical separation between nodes.

Among the different optically active quantum network platforms, trapped ions have consistently shown the most favorable rate-fidelity tradeoffs, with polarization encoding achieving the highest entanglement rates in ions \cite{stephenson2020high, oreilly2024fastphotonmediatedentanglementcontinuouslycooled}. Despite its advantages, polarization encoding suffers from uncontrolled polarization rotations in optical fibers, which directly impact entanglement fidelity after transmission. The severity of this polarization fluctuation varies substantially across fiber links. Previous work has shown that free drift in a quiet underground link can reduce the state fidelity by only a few percent over several hours \cite{Mika_thesis_paper}. In noisier environments, one minute of polarization instability can reduce process fidelity to below 80\% \cite{mckenzie2024clock, banner2026bifrost}. Where drift is significant, active polarization stabilization can maintain high-fidelity entanglement, but at the cost of periodically pausing entanglement generation for polarization correction at the photon wavelength \cite{banner2026bifrost}. Even in a comparatively quiet link, this correction was set to take approximately $1.1$ s, repeating every $100$ s \cite{kucera2024demonstration}. For noisier fibers, polarization drifts could significantly impact the duty cycle between entanglement generation and polarization stabilization.

Time-bin encoding of photonic qubits is robust against polarization instability, since the qubit states only differ in time of arrival. Node-photon or node-node time-bin protocols have been demonstrated across many matter-qubit platforms, such as color centers in diamond \cite{tchebotareva, knaut2024entanglement}, atomic ensembles \cite{farrera2018entanglement}, quantum dots \cite{jayakumar2014time, appel2022entangling}, single atoms \cite{li2025parallelized} and ions \cite{saha2025high}. Previous work in ions has demonstrated a fidelity of $0.970(4)$ at a sub-hertz entanglement generation rate,  via direct time-bin ion-photon entanglement generation \cite{saha2025high}. 
However, the entanglement generation sequence requires coherent manipulation of the ion state and two separate excitations for each entanglement attempt, limiting the attempt rate. 
Moreover, for any recoiling emitter platform, the direct time-bin scheme entangles the photon with the emitter's motional state, leading to decoherence and requiring mitigation strategies \cite{yu2026entanglement, apolin2026recoil}. 

Polarization-to-time-bin conversion using asymmetric interferometers circumvents both limitations of direct time-bin generation. In principle, the encoding stage can achieve entanglement rates similar to polarization encoding. It also enables heterogeneous networking between platforms with different native encodings \cite{nehra2026photonic}. Conversion between time-bin and polarization with entangled photon pairs is an established technique \cite{ photon_conversion1, photon_conversion2, photon_conversion3, photon_conversion4, nehra2026photonic}. To date, entanglement preserving time-bin-to-polarization conversion in matter-qubit-photon entangled states has been shown in a diamond color center \cite{vasconcelos2020scalable}, enabling polarization-basis measurements between the node and emitted photon. Classical correlations have been reported in quantum dots after the reverse encoding conversion \cite{yu2015two}. Here, we report the first entanglement-preserving polarization-to-time-bin qubit conversion in a matter-qubit-photon system, together with independent work by Haen \textit{et al}. \cite{haen2026telecom}.

In this work, entanglement is first generated between a $^{88}$Sr$^{+}$ ion and the polarization of an emitted $1092$~nm photon. The photon attenuation in silica fibers for this wavelength is low ($0.74$ dB/km) \cite{agrawal2012nonlinear}, and hence allows for polarization-to-time-bin conversion through an asymmetric Mach--Zehnder interferometer with low loss in the delay line ($< 0.044$ dB). Additionally, the low photon loss in standard telecom fibers (SMF-28) enables kilometer scale quantum networks directly at the emission wavelength  \cite{Mika_thesis_paper}. We demonstrate a time-bin state fidelity $>0.9$, and study the robustness of the converted state against a depolarizing channel. 

\section{Experimental Setup}\label{sec:exp}
\begin{figure*}
  \centering
  \includegraphics[width=\linewidth]{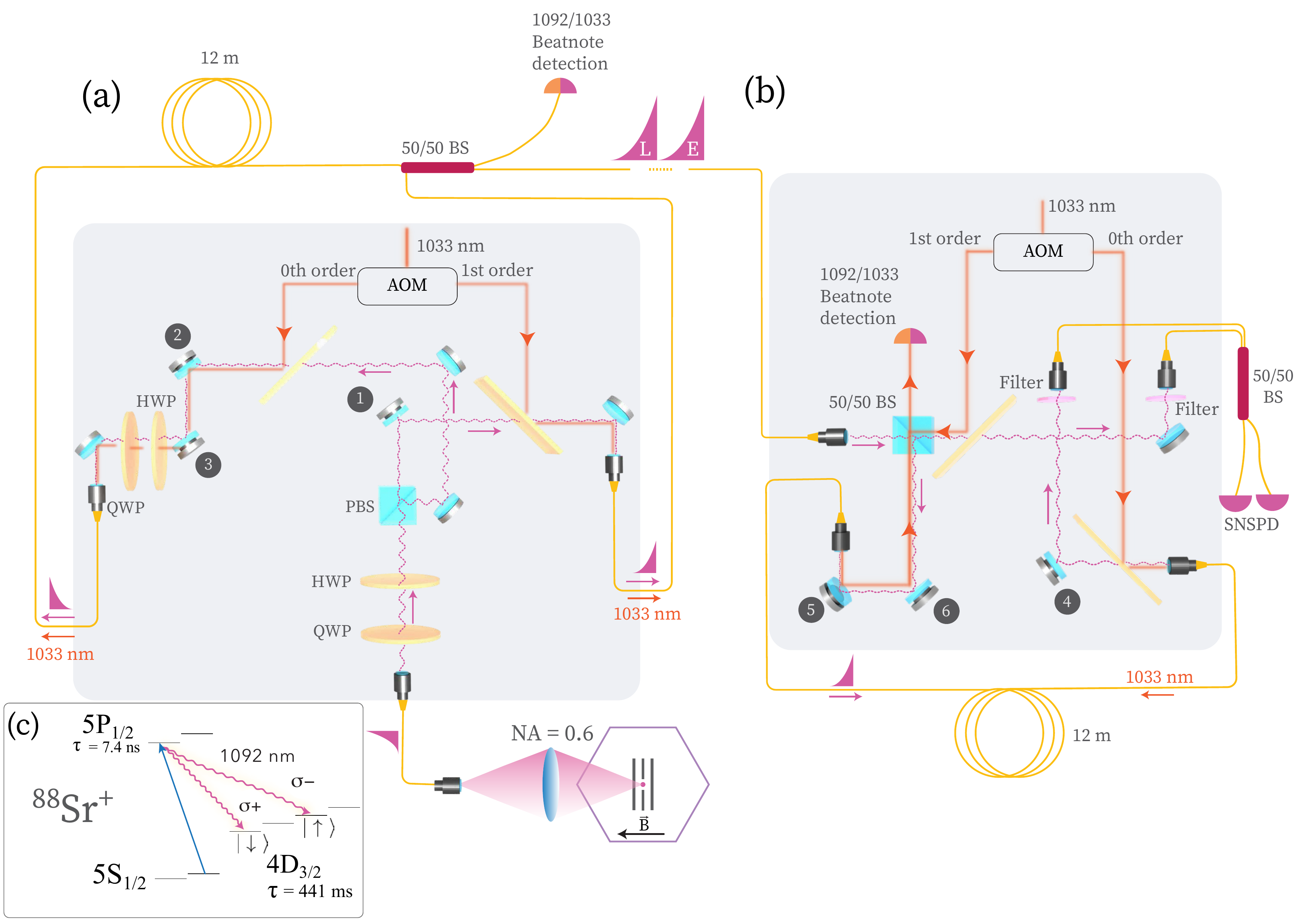}
  \caption{(a--b) Encoder and decoder asymmetric Mach--Zehnder interferometers for the time-bin encoder (a) and decoder (b). (a)~Waveplates set the polarization basis prior to conversion. A polarizing beam splitter separates the path of the horizontally and vertically polarized photons. The difference in length between the interferometer arms shifts the photons by $ 60$ ns in time relative to each other. Waveplates before the $12$ m delay line ensure the photons are identical in polarization degree of freedom when recombined at the 50/50 beam splitter, so the information is only encoded in time-bin. Piezos 2 and 3 are the $1033$~nm lock actuators. Piezo 1 corrects for drift between the photon and $1033$~nm lock paths every $30$~s. Full details of the phase stabilization scheme are given in Appendix~\ref{sec:phase_lock}. (b) Photon X-basis measurement interferometer. The phase stabilization scheme is identical to (a), with piezos 4--6 labeled accordingly. See text for details on the measurement. (c) Diagram of the $^{88}$Sr$^{+}$ ion levels involved in the ion-photon entanglement generation. The ion qubit states $\ket{\downarrow}$ and $\ket{\uparrow}$ are the metastable levels $\ket{4\text{D}_{3/2}, -3/2}$ and $\ket{4\text{D}_{3/2}, +1/2}$, respectively.} 
  \label{fig:setup}
\end{figure*}
To generate ion-photon entanglement, the $^{88}$Sr$^{+}$ ion is confined in a linear Paul trap with secular frequencies $(\omega_x, \omega_y, \omega_z) = 2\pi \times (1.23, 1.36, 0.15)$~MHz. A magnetic field of $4.3$~G along the photon collection axis lifts the degeneracy of the Zeeman states, shown in Fig.~\ref{fig:setup}(c). The qubit states $\ket{\downarrow}$ and $\ket{\uparrow}$ correspond to the Zeeman states $\ket{4\text{D}_{3/2}, -3/2}$ and $\ket{4\text{D}_{3/2}, +1/2}$, respectively. A custom objective with numerical aperture NA~$= 0.6$ collects the 1092~nm photons and couples them into an SMF-28 fiber \cite{Mika_thesis_paper}.

 The entanglement sequence starts by Doppler cooling the ion for $200$~$\mu$s. In each attempt of the entanglement generation loop, the ion is first prepared in $\ket{5\text{S}_{1/2}, +1/2}$ via optical pumping. A 20~ns pulse at 422~nm then excites the ion to $\ket{5\text{P}_{1/2}, -1/2}$, from which it decays with a $5.56(12)\%$ probability  \cite{barakhshan2021portal} along the 1092~nm channel, emitting a single photon. Attempts take 1.864 $\mu$s each and repeat up to 50 times or until a photon is detected, after which the sequence restarts \cite{Mika_thesis_paper}.
 
By collecting the photons along the quantization axis, we measure only $\sigma^+$ and $\sigma^-$ light, with ideally no $\pi$-light contribution~\cite{luo_pi_light}. Waveplates at the fiber output map these to $(H, V)$ polarization, yielding the ion-photon entangled state
\begin{equation}
    \ket{\Psi}_{\text{ion-photon}} = \frac{\sqrt{3}}{2} \ket{\downarrow} \ket{H} + \frac{1}{2} \ket{\uparrow} \ket{V}.
    \label{ent}
\end{equation}
The amplitude imbalance follows from the unequal Clebsch--Gordan coefficients of the two decay channels. A Bell state between two remote ions can still be generated from unbalanced ion-photon states at the cost of a reduced entanglement generation rate \cite{Unbalanced1, Unbalanced2}.

Photon detection events are accepted within a 20~ns window at a fixed delay after excitation. For time-bin-encoded photons, two such windows, early and late, are defined for Z-basis measurement. Detection on either superconducting nanowire single-photon detector (SNSPD) heralds entanglement and triggers ion state readout. For measurements in the $X$ basis, a Raman $\pi/2$ pulse detuned by 17.3~GHz from the 1004~nm transition rotates the ion qubit prior to shelving. Our shelving scheme detects $\ket{\downarrow}$ as a bright ion and all other states as dark, using $10\,\mu$s shelving pulses followed by an $8$~ms fluorescence detection window. The $\ket{\uparrow}$ population is measured in a separate experiment where a Raman $\pi$-pulse maps it to bright before shelving. We post-select on bright outcomes in both experiments, which rejects population that has leaked outside the qubit manifold. Full details of the readout and post-selection scheme are given in~\cite{Mika_thesis_paper}.
 
The photon basis conversion from polarization to time-bin happens in an encoder interferometer shown in Fig.~\ref{fig:setup}(a). A polarizing beam splitter (PBS) separates the vertically and horizontally polarized photons from the state in Eq.~\ref{ent} into distinct paths on an asymmetric Mach--Zehnder interferometer. Waveplates at the board input ensure the photon polarization is aligned to the PBS axes. The long interferometer arm has a $12$~m delay line, which transmits $>0.99$ of the $1092$~nm photons. Waveplates in the long arm ensure that the paths recombine with the same polarization on the 50/50 beam splitter (BS). This introduces an unavoidable loss of 50\% when using only passive optical elements. However, a combination of a PBS and an electro-optic modulator (EOM) could be used instead to substantially reduce this loss \cite{nehra2026photonic, photon_conversion3}. Photons at the output port of the beam splitter are identical in every degree of freedom except time of arrival. The board therefore converts the state to
\begin{equation}
    \ket{\tilde\Psi}_{\text{ion-photon}} = \frac{\sqrt{3}}{2} \ket{\downarrow} \ket{E} + \frac{e^{i \Delta \phi_e}}{2} \ket{\uparrow} \ket{L},
    \label{ent_tb}
\end{equation}
where $\ket{E}$ and $\ket{L}$ denote the early and late time-bins, respectively. The time-bins are separated by $60$~ns, which is $\approx 8\times$ the excited state lifetime of $7.4$~ns. $\Delta \phi_e$ is the optical phase difference between the two interferometer arms of the encoder board.


We bound the fidelities of the polarization and time-bin encoded ion-photon states as detailed in Appendix~\ref{sec:bounds}. This characterization requires measurements of the ion-photon state in the $Z \otimes Z$ and $X \otimes X$ bases. Rotations on the photon qubits depend on their encoding. Polarization qubits are easily manipulated with waveplates. Photon arrival time determines the Z-basis measurement for time-bin encoded photons. Their X-basis measurement requires a rotation to the diagonal time-bin basis. For this purpose, the photons travel through a second asymmetric Mach--Zehnder interferometer with the same path length difference, shown in Fig.~\ref{fig:setup}(b). Photons traveling through both interferometers can have three distinct arrival times. After the encoding interferometer, the photons arrive early or late. An early (late) photon taking the short (long) decoder arm arrives first (last). Early photons taking the long decoder path or late photons taking the short decoder path instead arrive together at the beam splitter in a middle time-bin. This occurs with probability $0.5$, and a photon detection on SNSPD~1 or SNSPD~2 projects the ion state into
\begin{equation}
    \ket{\Psi}_\text{ion} = \frac{\sqrt{3}}{2} \ket{\downarrow} \pm \frac{e^{i(\Delta \phi_{e} - \Delta \phi_{d})}}{2} \ket{\uparrow},
\end{equation}
where a measurement in SNSPD~1 corresponds to the positive sign and on SNSPD~2 to the negative sign. $\Delta \phi_{d}$ is the optical phase difference between the two arms of the decoder interferometer. Relative fluctuations in the optical path difference between arms in encoding and decoding interferometers directly affect the state fidelity. Both boards in Fig.~\ref{fig:setup} employ active phase stabilization, detailed in Appendix~\ref{sec:phase_lock}. Each board uses a dual piezo-mirror-based lock with $14$~kHz and $12$~kHz bandwidth for the encoder and decoder interferometers, respectively. Light at $1033$~nm is used to lock the interferometer. The $1033$~nm lock light is shuttered for $200$~ns during the photon detection window of each entanglement attempt, to avoid an increase in background photon counts. Because the optical paths of the $1033$~nm lock and the photon do not fully overlap, residual phase drift between them is not corrected by the fast lock alone. We pause entanglement attempts once every $30$~s and send $1092$~nm laser light co-propagating (counter-propagating) with the photon path on the encoder (decoder) board. A separate piezo-mirror in the photon path corrects this remaining drift.


\begin{figure}
    \centering
    \includegraphics[width=0.9\linewidth]{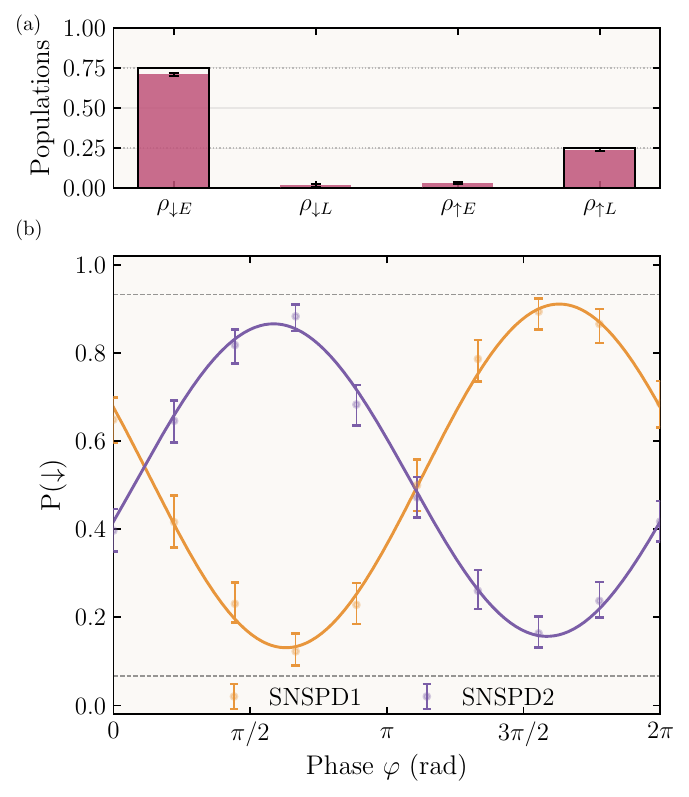}
    \caption{Fidelity of the time-bin encoded ion-photon entangled state. (a) Populations of the time-bin encoded ion-photon state. Each column shows a measured diagonal term of the density matrix. The black boxes indicate the ideal populations. (b) Diagonal basis correlations of the time-bin encoded ion-photon state. When the photons are in an early/late superposition, the probability of measuring a bright ion changes sinusoidally as a function of the phase of the ion analysis pulse. The dashed lines show the theoretical amplitude given our unbalanced ion-photon state. From (a) and (b), we calculate bounds on the converted ion-photon fidelity of $0.906 \pm 0.011 \le \mathcal{F} \le 0.934\pm 0.011$.}
    \label{fig:pop_and_coh}
\end{figure}

\section{Results}
\subsection{Fidelity bounds of the time-bin encoded ion-photon state}

To determine the ion-photon fidelity bounds after converting to time-bin encoding, we first measure the ion and photon in the $Z\otimes Z$ basis. The results of this measurement are the diagonal density matrix terms, i.e., the populations. Fig.~\ref{fig:pop_and_coh} shows the ion-photon state population measurements and their ideal values given the state Eq.~\ref{ent_tb}.  Next, we perform $X\otimes X$ measurements on the ion-photon state. To this end, the converted photons are sent to the decoder board. Prior to measuring the ion state, we apply a Raman $\pi/2$ pulse. Scanning the phase of this pulse yields the coherence fringe of Fig.~\ref{fig:pop_and_coh}(b), which shows the conditional probability of measuring state $\ket{\downarrow}$ given a detected photon on SNSPD channel 1 or 2. As a consequence of our unbalanced ion-photon state, the coherence contrast is limited to $C \le \sqrt{3}/2$, shown as dashed lines in Fig.~\ref{fig:pop_and_coh}(b). 

The time-bin encoded state fidelity is $0.906 \pm 0.011 \le \mathcal{F} \le 0.934\pm 0.011$. Fidelity bound measurements for the unconverted polarization ion-photon state give $0.9294 \pm 0.0075 \le \mathcal{F} \le 0.9598 \pm 0.0064$, which is an upper bound on what time-bin conversion can preserve. Errors in the polarization-to-time-bin conversion are listed in Table~\ref{tab:error_budget}. Appendix~\ref{sec:error_budget} details each error source. The main source of error is phase instability on the encoding and decoding interferometers. Phase errors are dominated by drift between the photon optical path and the 1033~nm fast phase lock, attributable to laboratory temperature drift and mechanical resonances of the interferometer breadboards within the laboratory vibration spectrum. We characterize the conversion's impact on the entanglement generation rate in Appendix~\ref{sec:rate}. 

\begin{table}[t] 
\centering
\caption{Error budget for the polarization-to-time-bin conversion. Each contribution is given as the resulting reduction to the ion-photon fidelity bounds, derived in Appendix~\ref{sec:bounds}.}
\begin{tabular}{lc}
\hline
Error Source & Fidelity Reduction \\
\hline
Interferometer Phase Instability & $< 0.022(1)$ \\
Temporal Mode Overlap & $0.0032(6)$ \\
Interferometer Arm Imbalance & $0.0014(4)$ \\
Background Photon Counts & $< 10^{-3}$ \\
\hline
Total & $< 0.028$ \\
\hline
\end{tabular}
\label{tab:error_budget}
\end{table}
\subsection{Robustness of converted state to polarization noise}
\begin{figure*}
  \centering
  \includegraphics[width=0.95\linewidth]{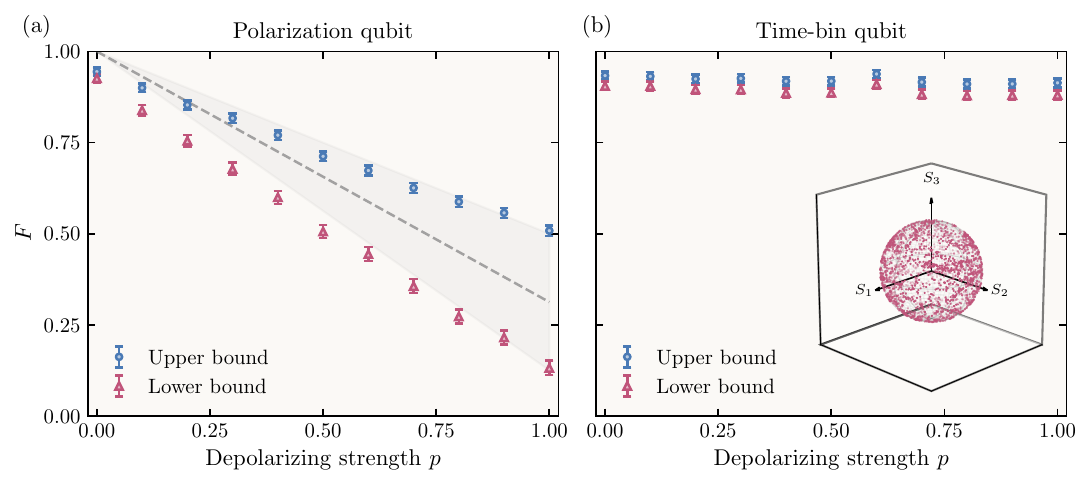}
  \caption{Fidelity bounds for the polarization (a) and time-bin (b) encoded ion-photon state as a function of the depolarizing strength $p$.
  While the polarization-encoded ion-photon state fidelity degrades as a function of $p$, the time-bin state fidelity shows no statistically significant degradation. Data at $p = 0$ and $p = 1$ are measured with the depolarizing channel off and on, respectively; intermediate points are statistical mixtures of the two datasets with weight $p$. Inset: Stokes parameters measured at the output of the fiber squeezer given a horizontally polarized input laser, recorded for one minute, showing depolarization at $p=1$. Each point in (a) and (b) takes over one minute to acquire. }
  \label{fig:fidelity_vs_p}
\end{figure*}
To demonstrate the robustness to polarization noise, we subject the polarization and time-bin encoded ion-photon states to a depolarizing channel. This channel is implemented using a fiber squeezer (PCD-M02/MPC, Luna), placed before the photon state analyzers. The implementation is detailed in Appendix~\ref{sec:depol_channel}. Fig.~\ref{fig:fidelity_vs_p} shows the ion-photon fidelity bounds for polarization and time-bin encoding as a function of depolarizing strength $p$. For $p = 0$, we measure the ion-photon state in the $Z \otimes Z$ and $X \otimes X$ bases with no noise applied. For $p = 1$, we repeat the measurement with the fiber squeezer on. For $0 < p < 1$, we draw from the two datasets with sampling probability $p$. The polarization fidelity decreases rapidly with $p$, and follows the theoretical expectation for the bounds given by the gray-filled region on the plot. At lower $p$ values, errors due to the imperfect initial ion-photon state dominate, and the experimental points differ from the expectation.  The time-bin fidelities at $p=0$ and $p=1$ agree to within 1.4 standard deviations, showing no statistically significant degradation and demonstrating that the time-bin state is robust against even the strongest polarization noise, shown in the inset of Fig.~\ref{fig:fidelity_vs_p}(b).



\section{Outlook}

 Our conversion method is far more robust to recoil-induced decoherence when compared to direct time-bin generation. For the polarization-to-time-bin conversion approach, only recoil decoherence due to random emission time differences remains. This effect is one to two orders of magnitude lower than the direct generation time-bin synchronization error \cite{yu2026entanglement}.  Therefore, mitigation strategies like trap-frequency synchronization or spin-dependent kick corrections \cite{yu2026entanglement, apolin2026recoil} are not necessary when converting from polarization to time-bin. 

The advantages of the conversion method over direct polarization encoding depend heavily on the specific fiber link. For quiet, underground networking fibers, polarization stabilization can be done infrequently and does not substantially affect the entanglement generation duty cycle. For fiber links with aerial sections or other disturbances, the polarization-stabilization downtime needed to match the converted state fidelity must be weighed against the interferometer drift correction downtime.

The conversion's infidelity, rate reduction, and phase-correction duty cycle are fixed overheads with clear paths for improvement. The costs are fixed by the local conversion apparatus and do not depend on the deployed fiber link. The approach generalizes to any qubit platform with polarization as its natural photon encoding, such as other trapped ion species and neutral atoms, and can in principle achieve entanglement rates comparable to polarization encoding. The dominant conversion error is interferometer phase instability. This error is specific to our current interferometer construction and lab environment, not a fundamental limit of the conversion technique itself, and can be reduced with improved mechanical and thermal isolation. Improved interferometer passive stability would allow for a lower drift-correction frequency, thus improving the experimental duty cycle between photon transmission and phase correction. The 50\% photon loss from the encoder board's passive beam splitter recombination could be recovered using a polarizing beam splitter, where an EOM would match the early and late photon polarizations \cite{nehra2026photonic, photon_conversion3}. With these improvements, the converted time-bin state fidelity and rate would match those of native polarization encoding, while the converted state remains robust to polarization noise in any fiber link, allowing for efficient deployment of hybrid quantum links across urban environments.

\begin{acknowledgments}
We thank Marko Cetina for design of the piezo-mirror mount, Ashish Kalakuntla for providing the time-bin entanglement gateware, and Michael Straus and Isabella Goetting for helpful discussions. This work was supported by the Army Research Office (grants W911NF-19-10296, W911NF-17-S-0002-0, W911NF-19-20181, and W911NF-22-10032), the National Science Foundation Convergence Accelerator program (OIA-2134891), and the Software-Tailored Architecture for Quantum CoDesign (STAQ) Award (PHY-2325080), as well as funding from Duke University under the Beyond-the-Horizon and DST-Launch initiatives.
\end{acknowledgments}

\bibliography{references}
\appendix
\section{Phase lock characterization}\label{sec:phase_lock}
\begin{figure*}
    \centering
    \includegraphics[width=1\linewidth]{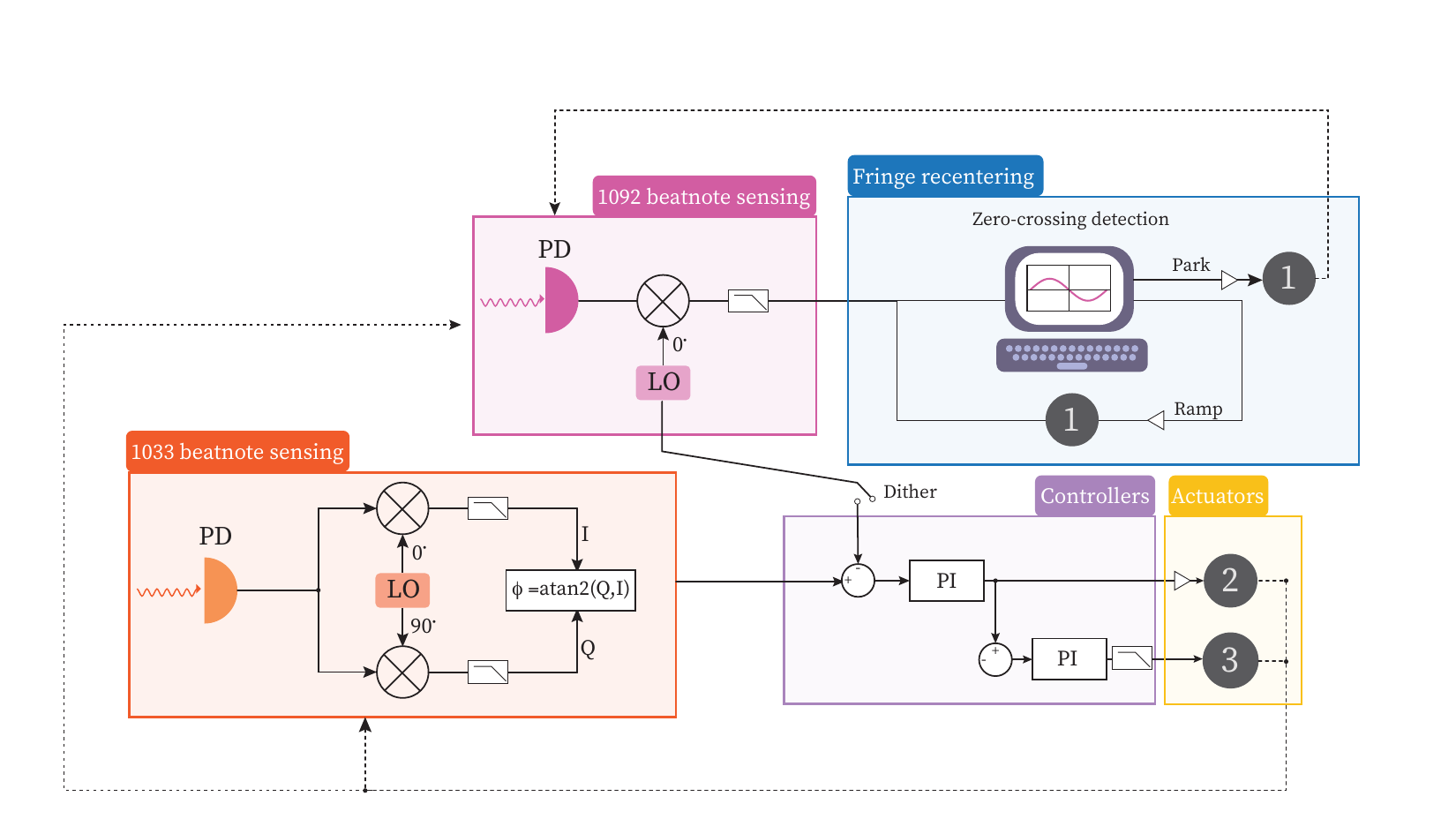}
    \caption{Block diagram of the complete phase stabilization scheme. The piezo numbers 1--3 are the ones shown in Fig.~\ref{fig:setup}(a) for the encoder board. The diagram is identical for the decoder board. The $1033$ nm lock is engaged continuously, but holds its PI value for $200$ ns during the photon detection time, when the stabilization beam is off. The beatnote is measured on a photodiode, and demodulated with analog mixer and low pass to  extract the in-phase ($I$) and quadrature ($Q$) components. From them, an FPGA calculates the phase, which is the error signal for feeding back to actuator 2. Actuator 3 is a long-range piezo responsible for the rail control of actuator 2. Piezo-mirrors 2 and 3 affect the optical phase measured in the $1033$~nm and $1092$~nm sensing paths. Piezo-mirror 1 only affects the $1092$~nm path. Arrows from actuators to sensing blocks schematically represent these connections and close the lock loops. Once per $30$ s, we send $1092$ nm laser light through the photon path to correct for slow drifts with a dither-based error signal generation. We ramp piezo 1 to find the fringe zero-crossing voltage and park it at the optimal position before returning to the ion-photon entanglement generation. }
    \label{fig:lock_electronics}
\end{figure*}
Fig.~\ref{fig:lock_electronics} shows the complete phase stabilization scheme, which is identical for the encoding and decoding interferometers. The stabilization combines a continuous fast lock using 1033~nm light with an intermittent fringe recentering at the photon wavelength (1092~nm), to correct for slow drifts not captured by the fast lock.

The 1033~nm portion of the scheme performs a continuous but partial stabilization of the interferometer, as the lock light does not co-propagate with the entangled photons throughout the entire path. It stabilizes most of the interferometer, including the 12~m delay line. However, the scheme relies on the passive stability of the non-stabilized portions. To this end, both interferometers sit on 110~mm thick honeycomb-core steel breadboards, which rest on Sorbothane sheets on an air-cushioned optical table. All cables going to the breadboards are clamped, and each board is enclosed in a box lined with acoustic damping foam, with rubber padding between the table and boxes to reduce the effect of air currents.

Refractive indices depend on both temperature and wavelength, so a large wavelength difference between lock and photon light could lead to non-common-mode phase drifts between the stabilization and photon paths. We therefore choose the 1033~nm lock wavelength due to its proximity to the photon wavelength (1092~nm) and the availability of a laser with 1~MHz linewidth, locked to a 1500-finesse cavity. However, the $15.7$~THz difference between photon and lock light precludes single-photon detection in the presence of lock light due to an increase in background counts that cannot be solved by additional filtering. The $40$~THz Raman gain spectrum in silica fibers \cite{agrawal2012nonlinear} means that 1033~nm light traveling through the delay line can transfer energy to vibrational modes of the glass and generate photons at 1092~nm via a Stokes Raman process, which cannot be separated from the entangled photons. Consequently, the lock light is shuttered by an AOM for $200$~ns every $1.864$~$\mu$s entangling attempt.

The 1033~nm beatnote is detected on a 400~MHz bandwidth photodetector. The interferometer's short and long paths are frequency-shifted relative to each other by an AOM, so the beatnote oscillates at the AOM drive frequency ($\sim$200~MHz). This RF signal is demodulated by mixing with the same reference used to drive the AOM, at 0\textdegree{} and 90\textdegree{} phase offsets, and low-pass filtered to extract the in-phase ($I$) and quadrature ($Q$) components. The interferometer phase is then computed on an FPGA as $\phi = \text{atan2}(Q, I)$, which is independent of optical power and interferometric visibility fluctuations. The custom FPGA gateware is a modification of Linien \cite{Linien}. When 1033~nm light is off for photon detection, both $I$ and $Q$ vanish, which the FPGA uses to detect the gating window and hold the error signal at zero, preventing the PI loop from responding. The interferometer phase is sufficiently stable over 200~ns that this sample-and-hold introduces negligible error.

The error signal is fed into a PI controller whose amplified output drives a piezo-mirror actuator. The piezo (PA25FE) has a high resonance frequency of $350$~kHz, allowing for fast feedback. Bandwidth is primarily limited by mechanical resonances of the piezo-mirror assembly. The piezo is glued to a tungsten carbide cylinder, with a 7~mm mirror glued to the piezo face. The assembly is held in place by a nylon casing around the cylinder, mounted in a standard mirror mount. The stiff tungsten carbide cylinder pushes mechanical resonances to higher frequencies, extending the loop bandwidth. The nylon casing decouples the piezo-mirror assembly from the mirror mount, to keep the high mechanical resonance frequency. The fast piezo's limited travel range (2.8~$\mu$m) is complemented by a second piezo (PK2FSF1) with a slower response but larger range (220~$\mu$m), which serves as rail protection for the first actuator. The combined closed-loop response achieves a bandwidth of 14~kHz (12~kHz) on the encoder (decoder) board, measured through a closed-loop sine sweep. 

To correct for drifts between the 1033~nm lock and photon optical paths, we implement a fringe recentering stabilization layer using 1092~nm light. In the experiment, we pause the ion-photon entanglement generation once per $30$ s, and use a 99\%/1\% beam splitter to send 1092~nm light co-propagating (counter-propagating) with the photons in the encoder (decoder) board. The third piezo, only present in the photon path, corrects the discrepancy between optical paths. To generate the error signal, we dither the setpoint of the fast lock at 2~kHz on the encoder and at 1.5~kHz on the decoder board. We find that dithering the setpoint is necessary to avoid thermal drift in the fringe recentering piezo. Dithering the recentering piezo directly heats it up, causing its voltage-to-displacement calibration to vary between shots. Moving the dither to the fast lock setpoint instead applies it to a closed-loop actuator, whose thermal drift is compensated by the servo, leaving the recentering piezo thermally stable. We choose dither frequencies 500~Hz apart to avoid cross-talk in the error signal generation, since the electronic paths are side by side. We measure the 1092~nm interference signal on a photodetector and mix it with the same reference used for the dither, before low-pass filtering it. After demodulation, the DC optical power term cancels, giving a signal proportional to $\cos(\phi)$ where $\phi$ is the interferometer phase. A voltage ramp on the recentering piezo linearly sweeps $\phi$, and generates the signal fringe. The signal is sent to an analog-to-digital converter, and then fitted to a cosine function to locate the zero crossing. A digital-to-analog converter then sets the third piezo voltage to the zero crossing point. This scheme is therefore independent of optical power and interferometer visibility fluctuations. The process takes $1.5$--$3$~s, and is currently limited by software overhead in the control system, but has room for optimization. During data acquisition, the recentering introduced a $4.8$--$9.1$\% overhead, corresponding to a duty cycle of $90.9$\%--$95.2$\% between photon transmission and phase correction, which was sufficient for this demonstration. The variation is due to the number of points taken for the recentering fit. 

\begin{figure}
    \centering
    \includegraphics[width=1\linewidth]{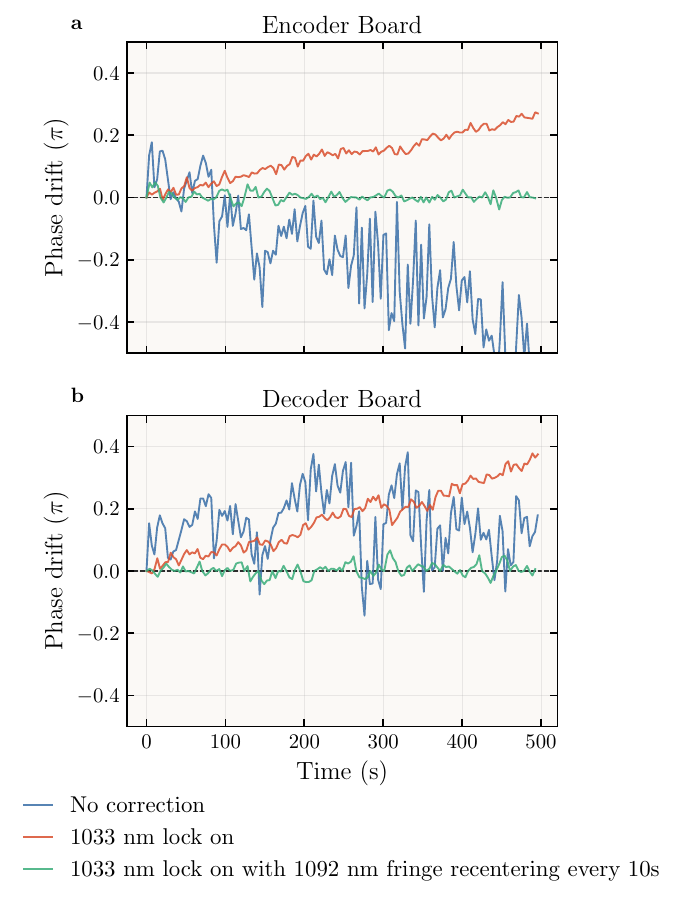}
    \caption{Phase lock performance of the (a) encoder and (b) decoder board over 500 s (8.3 min). With no correction, the phase freely fluctuates. With the $1033$ nm lock engaged, the phase improves substantially, but slow and monotonic drifts remain. The drift is mostly common mode between the two boards, which are next to each other on the same optical table. By correcting the photon path phase every $10$~s, the drift is controlled.}
    \label{fig:correction_performance}
\end{figure}

Fig. \ref{fig:correction_performance} shows the phase stabilization performance of each board during $500$ s on the photon path. When no correction is applied, the phase undergoes uncontrolled drift. The 1033 nm lock suppresses oscillations but has a slow monotonic drift. By recentering the interferometer fringe every 10~s, this drift is suppressed. In the experiment, we conservatively recenter once per $30$ s and observe no fidelity improvement from a higher recentering frequency. We attribute the tolerance to infrequent recentering to the largely common-mode drift between the two boards, which sit next to each other on the same optical table.
\section{Ion-photon fidelity bounds}\label{sec:bounds}
The ion-photon entanglement fidelity with respect to state Eq. \eqref{ent_tb} is
\begin{equation}
\begin{aligned} 
    \mathcal{F} &= \frac{3}{4} \rho_{\downarrow E, \downarrow E} + \frac{1}{4} \rho_{\uparrow L, \uparrow L}   + \frac{\sqrt{3}}{4} \rho_{\downarrow E, \uparrow L} + \frac{\sqrt{3}}{4} \rho_{\uparrow L, \downarrow E}.
\end{aligned}
\label{fidelity}
\end{equation}
The diagonal terms $\rho_{\downarrow E, \downarrow E}$ and $\rho_{\uparrow L, \uparrow L}$ can be directly measured from the encoder board, since they are the probabilities of measuring the photons in the early/late bin and the ion in the $\downarrow$/$\uparrow$ state. The off-diagonal terms $\rho_{\uparrow L, \downarrow E}$ and  $\rho_{\downarrow E, \uparrow L}$ cannot be directly measured, but can be bounded. An arbitrary single-qubit rotation is defined as
\begin{equation}
    R(\theta, \phi) = 
    \begin{pmatrix}
        \cos(\theta/2) & -i e^{-i \phi}\sin(\theta/2) \\
        -i e^{i \phi}\sin(\theta/2) &\cos(\theta/2)
    \end{pmatrix}
    \label{rotate}
\end{equation}
Our co-propagating 1004 nm Raman beams rotate the ion qubit. The angle $\theta = \Omega t$ is set by the Rabi frequency and pulse time, and the phase $\phi$ by the relative optical phase between the two Raman tones. Measuring the photon time-bin states in the $X$ basis requires rotating them to an equal superposition of $\ket{E}$ and $\ket{L}$. Given the input time-bin state with a time separation $\tau_{e}$, an asymmetric Mach--Zehnder interferometer with delay $\tau_{d} = \tau_{e}$ performs this rotation. For the photon, we are only able to perform $\phi = \pi/2$ rotations, which are sufficient for estimating upper and lower fidelity bounds. 
Defining the rotated density matrix
\begin{equation}
    \tilde{\rho} = R(\pi/2, -\pi/2)^{\otimes 2}\cdot \rho \cdot R^{\dagger}(\pi/2, -\pi/2)^{\otimes 2},
    \label{rotation}
\end{equation}
we may derive symmetric bounds on Eq. \eqref{fidelity} \cite{Auchter_2014},
\begin{equation}
    \mathcal{F_{-}} \le \mathcal{F} \le \mathcal{F}_{+},
\end{equation}
where
\begin{equation}
\begin{aligned}
    \mathcal{F}_{\pm} & = \frac{3}{4} \rho_{\downarrow E, \downarrow E} + \frac{1}{4} \rho_{\uparrow L, \uparrow L}  + \frac{\sqrt{3}}{4} (\tilde{\rho}_{\uparrow L, \uparrow L} +\tilde{\rho}_{\downarrow E, \downarrow E} \\ & - \tilde{\rho}_{\downarrow L, \downarrow L} - \tilde{\rho}_{\uparrow E, \uparrow E}  \pm 2 \sqrt{\rho_{\downarrow L, \downarrow L}  \rho_{\uparrow E, \uparrow E}} )
\end{aligned}
\label{lower_bound}
\end{equation}
The bounds are completely determined by diagonal elements of the original and rotated density matrices, all of which are directly measured. 
\section{Conversion error sources}\label{sec:error_budget}

\subsection{Interferometer phase stability}
\begin{figure*}
    \centering
    \includegraphics[width=1\linewidth]{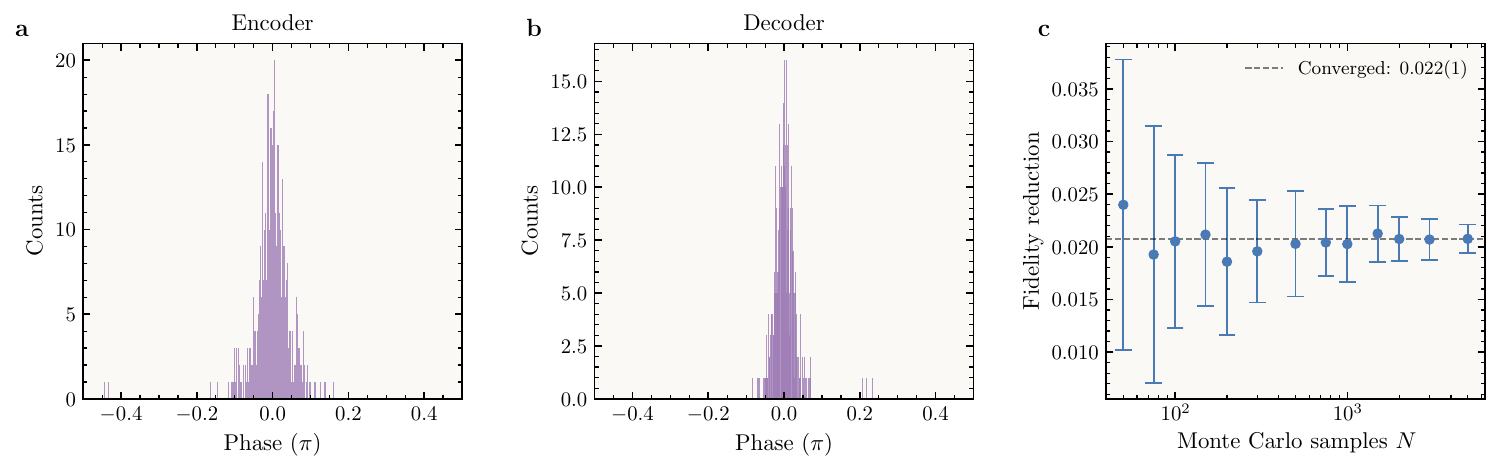}
    \caption{(a--b) Photon path optical phase data taken during the experimental run for the encoder and decoder boards. The phase on the encoder board is less stable due to a 14 Hz mechanical resonance of the breadboard-Sorbothane system. (c) Monte Carlo convergence plot showing the fidelity reduction due to phase instability versus number of Monte Carlo samples. For this simulation, phase was sampled from the distributions (a--b).}
    \label{fig:phase_distribution}
\end{figure*}

We collect data on the photon path phase stability on the encoder and decoder boards during experimental runs at every 1092 nm fringe recentering event. The encoder and decoder phase distributions are shown in Fig.~\ref{fig:phase_distribution}. The phase stability on the decoder board is noticeably better than on the encoder board. We attribute the difference to a 14~Hz mechanical resonance of the breadboard-Sorbothane system, identified by correlating the encoder phase oscillation frequency with the laboratory vibration spectrum measured by an accelerometer. Changing the Sorbothane feet thickness reduced but did not eliminate this resonance. A temperature fluctuation of $\sim 0.5\,^{\circ}$
C in the lab also contributes to the drift between paths. The mechanisms for this error are thermal expansion and thermo-optic drift in the portions of the path not shared between the photon and stabilization light, and the wavelength dependence of the thermo-optic coefficient causing non-common-mode optical path variation between the locked $1033$ nm light and the $1092$ nm photon in the delay lines.

Phase instability affects the ion-photon coherence measurements as
\begin{equation}
    \mathcal{C} = \frac{1}{2} + \frac{\sqrt{3}}{4} \cos(\phi + \Delta \phi_{e} - \Delta\phi_{d}) ,
\end{equation}
where $\phi$ is the phase of the Raman $\theta = \pi/2$ analysis pulse (Eq.~\ref{rotate}); $\Delta\phi_{e}$ and $\Delta\phi_{d}$ are the optical phase differences between the interferometer arms in the encoder and decoder boards respectively. By sampling $\Delta\phi_{e}$ and $\Delta\phi_{d}$ from the measured distributions in Fig.~\ref{fig:phase_distribution}(a--b), we can estimate the phase instability contribution to the polarization-to-time-bin conversion error. Since the two boards share common-mode drift and are positively correlated (see Fig.~\ref{fig:correction_performance}), independently sampling from their phase distributions overestimates the phase error, making this a conservative upper bound. A Monte Carlo simulation estimates it to be $< 0.022(1)$ on the upper and lower fidelity bounds, where Fig.~\ref{fig:phase_distribution}(c) shows the Monte Carlo convergence. 
\subsection{Temporal mode overlap}
\begin{figure}
    \centering
    \includegraphics[width=1\linewidth]{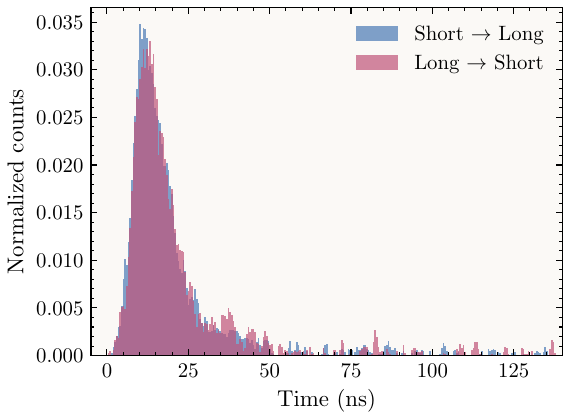}
    \caption{Temporal wavepackets from photons traversing the short (long) path on the encoder board and long (short) path in the decoder board. The X-basis time-bin measurement requires these distributions to be the same. We find a $\mathcal{V} = 0.9926(13)$ overlap.}
    \label{fig:temporal_overlap}
\end{figure}

 A mismatch in photon delays between interferometers leads to an erroneous $X$-basis photon measurement. To minimize the difference in delays $\Delta \tau = \tau_{e} - \tau_{d}$ between encoding and decoding interferometers, we measure the temporal wavepackets using a Time-Correlated Single Photon Counting (TCSPC) device, and we adjust the path length by splicing extra fiber into the encoder delay line. The photon wavepackets that travel through the short(long)/long(short) paths are shown in Fig.~\ref{fig:temporal_overlap}. We compute the overlap integral from the photon arrival histograms. The overlap is calculated from the raw data instead of a fit of the wavepacket form because the excitation AOM risetime ($\approx 20$~ns) is not negligible compared to the lifetime of the $5P_{1/2}$ $^{88}$Sr$^{+}$ excited state ($\tau = 7.4$~ns), leading to a distorted temporal envelope. The data are truncated at $4$ times the state lifetime to avoid integration over the low signal-to-noise tail. We define the overlap integral
\begin{equation}
    \mathcal{V} \equiv\int_0^{4\tau} \sqrt{p_{S,L}(t)\,p_{L,S}(t)}\;dt,
\end{equation}
where $p_{S,L}(t)$ and $p_{L,S}(t)$ are the probability densities of photon arrival time for a photon that traveled through the short encoder path and long decoder path, or vice versa. This equation gives an overlap of $\mathcal{V} = 0.9926(13)$ and a fidelity reduction of $0.0032(6)$ on both bounds.

\subsection{Interferometer arm imbalance}
Unbalanced losses in the interferometer arms affect the fidelity of the encoded state and of the coherence measurement. We measure and correct for losses with the SNSPD counts of a fluorescing $^{88}$Sr$^+$ ion driven on the 422~nm transition with 1033/1004~nm repumps, which provides background-free 1092~nm photon detection \cite{linke2012background}. By adjusting the kinematic mirror mounts on each board, we balance the arms to the optimal values shown in Table~\ref{tab:arm_balance}. The imbalances contribute a fidelity reduction of $0.0014(4)$.
\begin{table}[tb]
\caption{Ion fluorescence counts per second on SNSPD1 for photons going through different interferometer paths, when the photons traverse only one interferometer at a time.}
\centering
\begin{tabular}{lcc}
\hline
Board & Short path (cps) & Long path (cps) \\
\hline
Encoder & $463 \pm 18$ & $463 \pm 19$ \\
Decoder & $648 \pm 23$ & $639 \pm 21$ \\
\hline
\end{tabular}
\label{tab:arm_balance}
\end{table}

\subsection{Background photon counts}
Since the 1033 nm lock light is shuttered at the time of photon detection, we observe no increase of background photons due to the phase lock. Background counts are dominated by the SNSPD detector dark counts, at $\approx 15$ cps. In our previous work, they contributed only $< 10^{-3}$ infidelity \cite{Mika_thesis_paper}. In this work, the photon losses are higher as the photon travels through both encoder and decoder boards. As a consequence, the lower signal-to-noise ratio could contribute to higher ion-photon state infidelity. 

The ion-photon entanglement generation success probability can be used to estimate the effect of background detections on the measurements. A description of photon losses that affect the success probability can be found elsewhere \cite{Mika_thesis_paper}. 
In this experiment, we use a $20$ ns detection window for every measurement. On the day of data taking, we measure a success probability of $2.90(6)\times 10^{-4}$ for the unconverted ion-photon state, with polarization encoding. The value is the same for coherence and correlations measurements, since they require the same optical setup. For the converted time-bin state, we measure a success probability of $1.53(2)\times10^{-4}$ for ion-photon correlations. The lower value is due to recombining the photons on a 50/50 beam splitter on the encoder board, since only one output port is used. This is necessary for matching the time-bin polarizations with only passive optical elements. Losses on the polarization analysis board and time-bin encoder are similar, so that the time-bin correlations success probability is 52.8$\%$ of the polarization one. For the time-bin coherence measurements, we measure a success probability of $1.82(4)\times10^{-5}$. This value is lower than the one for correlations because of the additional photon losses on the decoder board and the $0.5$ probability of detecting photons in the middle time-bin. 

The shelving scheme employed in this work only detects a bright ion when its state is $ \ket{4\text{D}_{3/2}, -3/2}$ ($\ket{\downarrow}$). Therefore, at the end of an entangling attempt, a background photon detection results in a dark ion readout in more than 95\% of cases. The post-selection scheme discussed in Sec.~\ref{sec:exp} only accepts bright detection events. Consequently, only background events where the ion is also measured as bright affect the post-selected fidelity. For a 20 ns detection window, the probability of such events is $\sim 10^{-8}$, orders of magnitude below the entanglement generation success probabilities. Therefore, despite the additional loss of the photon basis conversion, the error due to background counts remains $< 10^{-3}$ for the time-bin ion-photon state.

\section{Entanglement generation rate}\label{sec:rate}
The entanglement generation rate for time-bin correlation measurements was  $26.1$ s$^{-1}$, including a $31.8$\% duty cycle due to pauses for cooling the ion. For time-bin coherence measurements, it was $3.10$ s$^{-1}$. 
For the unconverted polarization state, we measure an entanglement rate of $49.5$ s$^{-1}$ for both coherences and correlations. This value is lower than the one reported in \cite{Mika_thesis_paper} because it includes ion cooling time, and contains a higher polarization error in the optical pumping beam used for the entanglement generation, which primarily limits the achievable success probability. 

\section{Depolarizing channel characterization} \label{sec:depol_channel}
We implemented the depolarizing noise using a fiber squeezer (PCD-M02/MPC, Luna). This device can be modeled as four waveplates arranged in two pairs, each pair rotating the polarization around orthogonal axes on the Poincaré sphere. The combination of rotations is sufficient to transform any input polarization state to any output state.
An arbitrary $SU(2)$ rotation is, however, not enough to construct depolarizing noise. The transformation must also be uniformly random. A three-waveplate scheme uniformly scrambles the polarization~\cite{uniform_pol}. If $\theta_{1}, \theta_{2}$ and $\theta_{3}$ are the phase retardations of the first three fiber squeezer sections, then a random polarization state can be achieved by sampling $\theta_{1}$ and $\theta_{3}$ uniformly between $[-\pi, \pi]$ and $\cos(\theta_{2})$ uniformly between $[-1, 1]$. In practice, since the phase retardations are proportional to the applied voltages, we drive $\theta_{1}$ and $\theta_{3}$ with ramps and $\theta_{2}$ with an $\arccos$ function. A third ramp applied to $\theta_{4}$ mitigates imperfect orthogonality on the squeezer axes. Table~\ref{tab:squeezer} shows the drive parameters applied to each squeezer section.

Even though the rotations are deterministic, they  look effectively random if the scrambling period is much smaller than the integration time of the ion-photon experiment. Given that each point in our experiment (Fig. \ref{fig:fidelity_vs_p}) takes more than one minute, we verify this assumption by measuring the Mueller matrix of the scrambling transformation over one minute. For this measurement, we prepare the input 1092~nm laser light in horizontal, diagonal and circular polarization states, and measure the output polarization after the fiber squeezer on a polarimeter (PAX1000IR1, Thorlabs). 

\begin{figure}
    \centering
    \includegraphics[width=1\linewidth]{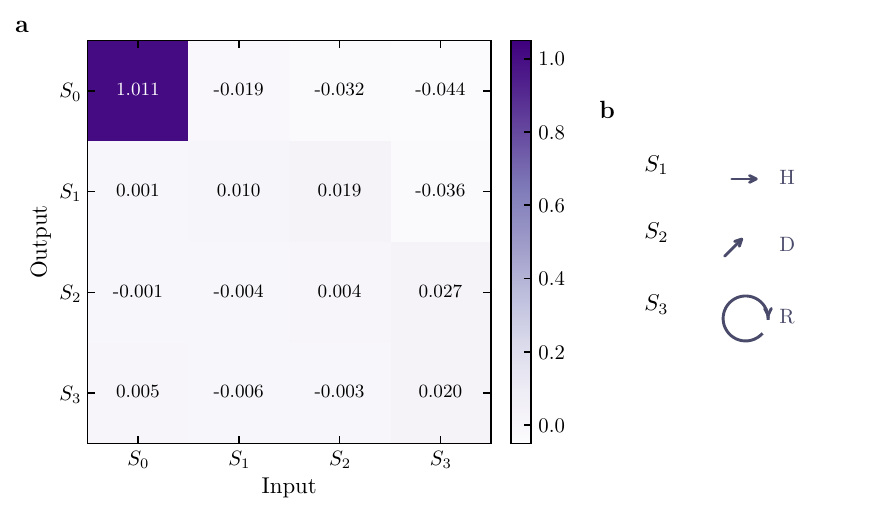}
    \caption{(a) Mueller matrix measurement of the depolarizing noise. An ideal depolarizer should have $M_{S_{0}, S_{0}}= 1 $, while all other entries vanish. $S_{i}$ are the Stokes parameters that define a polarization state. (b) Schematic representation of the Stokes parameter basis. $S_{1}$ corresponds to horizontal polarization, $S_{2}$ to diagonal polarization and $S_{3}$ to right-circular polarization.}
    \label{fig:mueller}
\end{figure}
Fig.~\ref{fig:mueller} shows the Mueller matrix. An ideal depolarizer does not introduce any attenuation, such that the first diagonal entry of the Mueller matrix $M_{S_{0}, S_{0}} = 1$. It should also destroy polarization structure, such that $M_{S_i,S_j} = 0$ for $i, j \ne 0$. The first element in the measured matrix is $M_{S_{0},S_{0}} > 1$, which is a nonphysical artifact of laser power fluctuations during data acquisition. The measured matrix also differs from the ideal depolarizer in its anisotropy, with normalized diagonal elements $\text{diag}(M) = \{1, 1-p_{S_1}, 1-p_{S_2}, 1-p_{S_3}\}$, normalized by $M_{S_{0},S_{0}}$, where $p_{S_1} = 99.01\%$, $p_{S_2} = 99.57\%$ and $p_{S_3} = 97.98\%$. The channel therefore has an average depolarization of $98.85\%$, sufficiently close to an ideal depolarizer for the purposes of this demonstration.

\begin{table}[tb]
\centering
\caption{Drive parameters for the four fiber squeezer channels used to implement the depolarizing channel.}
\begin{tabular}{ccccc}
\hline
Channel & Waveform & Frequency & $V_\text{pp}$ (V) & $V_\text{dc}$ (V) \\
\hline
$\theta_1$ & Ramp        & 101~mHz & 1.350 & 3.5 \\
$\theta_2$ & $\arccos$   & 11~Hz   & 1.350 & 3.7 \\
$\theta_3$ & Ramp        & 23~Hz   & 1.350 & 3.5 \\
$\theta_4$ & Ramp        & 67~Hz   & 1.350 & 3.5 \\
\hline
\end{tabular}
\label{tab:squeezer}
\end{table}
The action of this channel on the ion-photon density matrix is 
\begin{equation}
    \mathcal{N}(\rho) = \sum_{i = 0}^{3} (\mathds{1} \otimes K_{i}) \rho  (\mathds{1} \otimes K_{i})^{\dagger},
\end{equation}
where the Kraus operators $K_{0} = \sqrt{1-3p/4} \  \mathds{1}$, $K_{1} = \sqrt{p/4} \ X$, $K_{2} = \sqrt{p/4} \ Y$,  and $K_{3} = \sqrt{p/4} \  Z$ only act on the photon qubit. Here, $p = 0$ corresponds to the identity channel and $p = 1$ to a completely depolarized photon state.

\end{document}